# Assessing novice programmers' perception of ChatGPT: performance, risk, decision-making, and intentions

**John Paul P. Miranda, Jaymark A. Yambao**
College of Computing Studies, Mexico Campus, Don Honorio Ventura State University, Mexico, Philippines

| Article Info | ABSTRACT |
|---|---|
| *Article history:*<br><br>Received May 22, 2024<br>Revised Sep 18, 2024<br>Accepted Mar 18, 2025<br><br>*Keywords:*<br><br>AI in education<br>Digital skill development<br>Educational technology<br>Innovative learning tools<br>Programming students | This study explores the novice programmers' intention to use chat generative pretrained transformer (ChatGPT) for programming tasks with emphasis on performance expectancy (PE), risk-reward appraisal (RRA), and decision-making (DM). Utilizing partial least squares structural equation modeling (PLS-SEM) and a sample of 413 novice programmers, the analysis demonstrates that higher PE of ChatGPT is positively correlated with improved DM in programming tasks. Novice programmers view ChatGPT as a tool that enhances their learning and skill development. Additionally, novice programmers that have a favorable RRA of ChatGPT tend to make more confident and effective decisions, acknowledging potential risks but recognizing that benefits such as quick problem-solving and learning new techniques outweigh these risks. Moreover, a positive perception of ChatGPT's role in DM significantly increases the inclination to use the tool for programming tasks. These results highlight the critical roles of perceived capabilities, risk assessment, and positive DM experiences in promoting the adoption of artificial intelligence (AI) tools in programming education. |

*Corresponding Author:*

John Paul P. Miranda
College of Computing Studies, Mexico Campus, Don Honorio Ventura State University
San Juan, Mexico, Pampanga, Philippines
Email: jppmiranda@dhvsu.edu.ph

## 1. INTRODUCTION

In this fast-paced world of programming education, the integration of advanced tools and technologies has become essential for enhancing learning outcomes. One such tool, Chat Generative Pretrained Transformer (ChatGPT), has captured the attention of educators and students alike for its potential to assist novice programmers in programming-related tasks. Understanding how individuals learn their first programming language is a central focus of computing education research. Novice programmers, who are just beginning their journey in computer programming, often display characteristics that set them apart from more experienced coders. They typically have concrete, low-level, syntax-based knowledge acquired through introductory programming classes and often struggle with writing code proficiently, focusing mainly on a surface-level understanding of programs. These beginners tend to use a "line-by-line" approach to programming and have less developed mental models of computer programs compared to expert programmers, who possess more abstract representations [1]–[3].

Novice programmers face various challenges, such as difficulties in understanding programming language syntax and abstract concepts like objects and classes in object-oriented programming. They commonly make errors in their code and find it challenging to write code that leverages advanced concepts like parallelism and heterogeneity, typically expected from more experienced programmers [4]. Research has emphasized the importance of providing targeted examples for novice programmers, as example programs have been found to be highly beneficial for their learning process [5]. Novice programmers often face challenges in acquiring essential skills such as problem-solving, program design, comprehension, and debugging [6]. Despite these difficulties, many novice programmers still actively seek resources like documentation and code samples to enhance their understanding and practical skills [7]. In addition to this, researchers have been focusing on understanding the mistakes novice programmers make to improve the quality of programming education. Novice programmers tend to start writing programs before planning them, highlighting the importance of teaching effective planning strategies [8]. The field of novice programming has been extensively researched, with efforts to understand the main issues faced by these learners [9]. Novice programmers are often described as having 'fragile' knowledge, emphasizing the need for comprehensive educational approaches [10]. Metacognitive skills are essential for novice programmers to solve unfamiliar problems effectively [11]. Additionally, studies have explored the challenges novice programmers encounter while programming and their reactions during programming tasks [12]. Understanding the pain points of novice programmers in developing smart systems is crucial for improving their learning experiences [13].

Understanding how novice programmers perceive artificial intelligence (AI) tools is essential for improving tool design and usability, enhancing educational effectiveness, and fostering broader adoption and integration. Insights from beginners can help identify gaps and challenges, promote inclusive technology, and drive innovation and future development. Positive experiences with AI tools boost user confidence and empower novices to tackle complex tasks, contributing to a more skilled workforce and broader economic benefits. Insights into how beginners interact with AI tools can inform educators about the strengths and weaknesses of these tools in a learning environment. This understanding can help in integrating AI tools into programming curricula in a way that enhances learning outcomes. The perceptions of novice programmers can influence the broader adoption of AI tools in educational and professional settings. If beginners find these tools helpful and easy to use, they are more likely to continue using them and recommend them to others.

OpenAI's ChatGPT has shown significant potential in transforming various industries, including education [14]. Research indicates that integrating ChatGPT in educational settings can enhance idea generation, topic verification, proofreading, and editing, leading to positive student experiences, and perceptions [15]. Virtual tutors powered by ChatGPT can make the learning process more engaging for both students and teachers [16]. ChatGPT has been acknowledged as a valuable tool for simplifying complex concepts in health education [17] and as a supporting tool for academics [18], [19]. However, challenges and associated risks have also been noted [18], [20]–[25], with ChatGPT helping to alleviate the workload in various routine tasks in academia [23], [26], [27]. AI tools like ChatGPT can assist novice programmers in various ways [28]–[31]. These tools can help address challenges faced by beginners, such as issues related to basic program design, algorithmic complexity, and the fragility of novice knowledge. Utilizing ChatGPT for educational support can aid in tasks like program comprehension and improving their understanding of code execution [32]–[35]. ChatGPT's conversational and programming abilities make it an attractive tool for facilitating education for beginners [36]. Research has shown that ChatGPT can benefit beginner and intermediate programming courses by offering valuable guidance for both teachers and students in understanding and optimizing their solutions [37], [38]. The use of AI tools like ChatGPT can also benefit novice programmers when trying to understand small programs or exploring linguistic features in a new programming language.

This study investigates the perceptions of first-year undergraduate programming students (i.e., novice programmers) regarding ChatGPT, focusing on four key constructs: performance expectancy (PE), risk-reward appraisal (RRA), decision-making (DM), and intention to use (IU). AI tools like ChatGPT have the potential to significantly support novice programmers by providing educational assistance, aiding in program comprehension, offering guidance in solving programming exercises, and facilitating the learning process in various programming domains. Furthermore, this study explores how PE, RRA, and DM affect novice programmers' IU ChatGPT for programming-related tasks through a partial least squares structural equation modeling (PLS-SEM) analysis.

## 2. THEORETICAL BASIS
### 2.1. Performance expectancy

PE within the context of this study, refers to the anticipation that novice programming students have regarding the potential benefits of using ChatGPT to enhance their programming capabilities. This concept is derived from the unified theory of acceptance and use of technology (UTAUT) model, which identifies PE as a crucial factor influencing technology adoption and usage intentions. In programming, PE refers to the belief that using ChatGPT can lead to improved code quality, enhanced problem-solving efficiency, and the acquisition of new programming skills. Extensive research in the field of technology acceptance supports the notion that higher PE correlates with increased IU and actual use of technology [39]–[43]. When applied to ChatGPT, this means that if students perceive substantial benefits from using this tool for their programming tasks, they are more likely to rely on it, thereby enhancing their DM processes. Increased PE in ChatGPT can lead to better utilization of ChatGPT's capabilities, where students will use ChatGPT's features for complex problem-solving which can result in higher-quality programming outcomes. Additionally, with a strong belief in ChatGPT's ability to improve their programming tasks, students can make more informed and confident decisions during the coding process. This may lead to achieving better results in their programming assignments and projects. Therefore, the hypothesis is that increased PE in ChatGPT is associated with improved DM regarding programming tasks. This connection highlights the importance of fostering positive perceptions of ChatGPT's capabilities among novice programmers to enhance their educational experiences and outcomes.
*H1. Increased PE in ChatGPT is associated with improved DM with regards to programming tasks.*

## 2.2. Risk-reward appraisals

RRAs involves evaluating the potential benefits and risks of using ChatGPT in programming tasks. This concept is based on the expectancy theory of motivation, which suggests that individuals assess the value of specific actions based on their expected outcomes. For novice programmers, this appraisal includes weighing the accuracy and reliability of ChatGPT's programming solutions against potential risks such as developing overreliance or encountering misinformation. When deciding to use ChatGPT, programmers engage in a cognitive evaluation, considering whether the benefits (e.g., enhanced efficiency, better problem-solving) outweigh the risks (e.g., incorrect code suggestions, misinterpretation of programming concepts). A favorable RRAa is expected to enhance DM in programming tasks. If users perceive the utility of ChatGPT-such as its ability to improve programming outcomes and streamline workflows-as outweighing potential drawbacks, they are more likely to rely on it for assistance. This reliance, driven by a positive assessment of ChatGPT's capabilities which can leads to more effective and confident DM in programming tasks. This hypothesis posits that when programmers find the risk-reward balance of using ChatGPT to be favorable, they will experience improved DM capabilities, thereby enhancing their overall programming performance. *H2. A favorable RRAs of ChatGPT is linked to enhanced DM with regards to programming tasks.*

## 2.3. Decision-making

DM from the perspective of novice programmers, involves selecting among various alternatives to solve programming challenges, guided by the information and recommendations provided by ChatGPT. This process is influenced by cognitive biases, such as confirmation bias and overconfidence, which can shape the evaluation of suggestions provided by ChatGPT. A positive perception of ChatGPT's role in DM can mitigate some of these cognitive biases by offering a reliable source of information. This reliability encourages a more analytical and reflective approach to problem-solving, promoting better DM practices [24], [31], [33], [44]–[47]. The theory of planned behavior supports this hypothesis by suggesting that positive attitudes towards a behavior (using ChatGPT) are linked to stronger behavioral intentions-in this case, the inclination to use ChatGPT for programming tasks. When novice programmers view ChatGPT positively in the context of DM, they are more likely to rely on it as a valuable tool, enhancing their inclination to use it for solving programming problems. This positive attitude not only increases the likelihood of utilizing ChatGPT but also fosters confidence and effectiveness in their programming endeavors. *H3. A positive view on the role of ChatGPT in DM is connected to a greater inclination to use it for programming tasks.*

## 3. METHOD

This study adapted the proposed model in Figure 1 previously developed by Shahsavar and Choudhury [48]. The framework has four key constructs: PE, RRA, DM, and IU. Specifically, this research aims to understand the relationships among these constructs in the context of utilizing ChatGPT for programming tasks. In particular, the study examined how increased PE of ChatGPT is associated with enhanced DM regarding programming tasks, how favorable RRAs of ChatGPT are linked to improved DM, and how positive perceptions of role of ChatGPT in DM correlate with a greater inclination to use it for programming tasks. To rigorously analyze these relationships, PLS-SEM is utilized. This robust analytical approach aids in more in depth understanding how PE, RRAs, and DM processes interact to influence users' intentions to use ChatGPT.

### 3.1. Sample characteristics

As indicated in Table 1, of 1,134 individuals invited to participate in the survey, only 413 respondents were found eligible for analysis. This attrition rate (63.58%) suggests that while these respondents had previously used ChatGPT, they did not utilize it for programming-related tasks. The age range of the sample was from 17 to approximately 21 years old. The majority (97.3%) were enrolled in public schools. Gender distribution in the sample was predominantly male (76%), followed by female (21.8%), those who preferred not to specify their gender (1.9%), and non-binary individuals (0.2%).

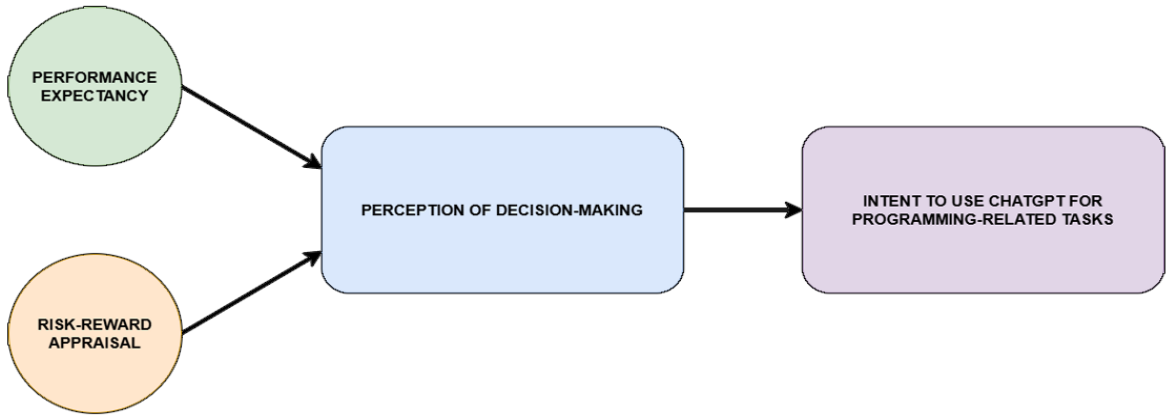

Figure 1. Proposed model based on Shahsavar and Choudhury [48]

Table 1. Description of the respondents

| Demographics | Frequency | Percentage |
|---|---|---|
| Age (mean±SD) | | 18.99±1.96 |
| School type | | |
| Public | 402 | 97.3 |
| Private | 11 | 2.7 |
| Gender | | |
| Male | 314 | 76 |
| Female | 90 | 21.8 |
| Non-binary | 1 | 0.2 |
| Prefer not to say | 8 | 1.9 |
| Academic program | | |
| Information technology | 207 | 50.1 |
| Computer science | 106 | 25.7 |
| Information system | 39 | 9.4 |
| Others | 61 | 14.8 |
| Preferred learning style | | |
| Visual | 156 | 37.8 |
| Auditory | 57 | 13.8 |
| Kinesthetic | 113 | 27.4 |
| Read/ write | 87 | 21.1 |
| Average weekly time spent on programming-related tasks outside of classes (mean±SD) | | 4.5 ± 5.239 |
| Primary operating system for programming tasks | | |
| Windows | 398 | 96.4 |
| macOS | 7 | 1.7 |
| Linux | 8 | 1.9 |
| Previous experience with generative AI tools other than ChatGPT | | |
| Yes | 249 | 39.7 |
| No | 164 | 60.3 |
| Frequency of using ChatGPT for programming tasks | | |
| Rarely | 208 | 50.4 |
| Occasionally | 157 | 38 |
| Frequently | 48 | 11.6 |
| Specific purposes for using ChatGPT in programming-related tasks* | | |
| Generating code snippets | 100 | 24.2 |
| Providing programming advice or recommendations | 210 | 50.8 |
| Debugging assistance | 98 | 23.7 |
| Explaining programming concepts | 251 | 60.8 |
| Others | 22 | 5.33 |
| Familiarity with ChatGPT's full capabilities | | |
| Slightly familiar | 156 | 37.8 |
| Somewhat familiar | 131 | 31.7 |
| Moderately familiar | 86 | 20.8 |
| Very familiar | 40 | 9.7 |
| Extent to which they find ChatGPT helpful in programming tasks | | |
| Slightly persuasive | 125 | 30.3 |
| Somewhat persuasive | 153 | 37 |
| Moderately persuasive | 97 | 23.5 |
| Very persuasive | 38 | 9.2 |
| Total | 413 | 100 |

Educationally, half of the respondents (50.1%) were pursuing a bachelor of science in information technology. Other fields of study included computer science (25.7%), information systems (9.4%), and various other disciplines like associate in computer technology, computer engineering, data science, and analytics (14.8%). The diverse educational backgrounds and learning preferences reflect the wide range of applications and approaches to integrating AI tools like ChatGPT in programming education. In terms of learning preferences, 37.8% favored visual aids like images, charts, and diagrams. Other learning styles included a hands-on approach (27.4%), reading and writing (21.1%), and auditory learning (13.8%). On average, the respondents used ChatGPT for programming tasks outside of class for about less than an hour to nine hours per week. Most participants (96.4%) used Windows for these tasks, followed by Linux (1.9%) and macOS (1.7%). A majority (60.3%) had no prior experience using generative AI tools other than ChatGPT. Usage frequency of ChatGPT for programming tasks varied: 50.4% rarely used it, 38% occasionally, and only 11.6% frequently.

As regards the purpose of using ChatGPT, over half (60.8%) utilized it for explaining programming concepts, while 50.8% sought programming advice. Approximately a quarter used it for debugging assistance (23.7%) and generating code snippets (24.2%), with a smaller fraction (5.33%) using it for academic tasks like essay writing. Familiarity with ChatGPT's full capabilities also varied: 37.8% were slightly familiar, 31.7% somewhat familiar, 20.8% moderately familiar, and only 9.7% were very familiar. In evaluating the tool's effectiveness, 37% found ChatGPT slightly persuasive in programming tasks, 30.3% moderately persuasive, 23.5% slightly persuasive, and 9.2% very persuasive.

## 3.2. Sampling and data collection

A non-probability snowball purposive sampling technique was used to obtain responses using Google Form. The target audience for this study were novice programmers (i.e., first year undergraduate students with programming-related subjects). Novice programmers in this study refer to undergraduate students currently enrolled in introductory programming courses [49]. At the same time, they should have prior experience, or have used ChatGPT at least once for programming-related tasks. These were the minimum requirements to participate in the study. The two were followed in order to increase the number of potential respondents for the study. For any successful recruit, they were asked to refer or send the survey form link to their classmates. The study utilized Google Form to survey respondents across universities. A consent in accordance with the existing data privacy laws in the country was also acquired from the respondents prior to the start of the survey.

## 3.3. Research instrument

The study applied the framework proposed by Shahsavar and Choudhury [48] and adapted it to the context of programming tasks for novice programmers. Three experts in educational technology and programming validated the instruments. To assess the internal consistency of the instrument, a pilot test was conducted with 93 respondents. Their data was excluded from the main PLS-SEM analysis which involved 413 actual respondents. Table 1 presents the Cronbach's Alpha values for the four constructs: PE (3 items, $\alpha=0.884$), RRA (3 items, $\alpha=0.793$), DM (3 items, $\alpha=0.831$), and IU (2 items, $\alpha=0.750$). The overall cronbach alpha of the instrument is 0.929. These values indicate good internal consistency which indicates that the items within each construct measure a single underlying concept reliably. This tool comprised two sections: one gathers personal information about novice programmers, including age, school type, gender, academic program, preferred learning style, weekly time spent on programming tasks outside of classes, primary operating system for programming tasks, previous experience with generative AI tools other than ChatGPT, frequency of using ChatGPT for programming tasks, specific purposes for using ChatGPT in programming-related tasks, familiarity with ChatGPT's capabilities, and the extent to which they find ChatGPT helpful in programming-related tasks. The second section assessed PE, RRA, DM, and IU, consisting of 11 statements to which respondents express their level of agreement (ranging from strongly disagree to strongly agree).

## 4. RESULTS AND DISCUSSION

This study utilized PLS-SEM approach to investigate the hypotheses using the proposed model in Figure 2. In Table 2, it shows that all variable items have factor loadings (FL) greater than 0.60, indicating strong correlations with their respective factors. This confirms that each item effectively measures its intended construct. Additionally, the average variance extracted (AVE) for each construct exceeds 0.50. This signifies that more than half of the variance in the indicators is captured by the construct which also confirms good convergent validity. Convergent validity is achieved when multiple items intended to measure the same construct do so effectively, as evidenced by high FL and sufficient AVE values. Table 3 further demonstrates this by comparing the square root of the AVE for each construct with the correlations between the constructs. Based to the Fornell-Larcker criterion, all constructs exhibit good discriminant validity; the square roots of the AVEs for DM, IU, PE, and RRA (0.826, 0.899, 0.855, 0.846, respectively) are greater than their correlations with other constructs (ranging from 0.48 to 0.662). This indicates that each construct is distinct and uniquely measures its respective aspect of the model. The model in this study produced a goodness of fit (GoF) value of 0.579 which indicates an adequate fit of the model.

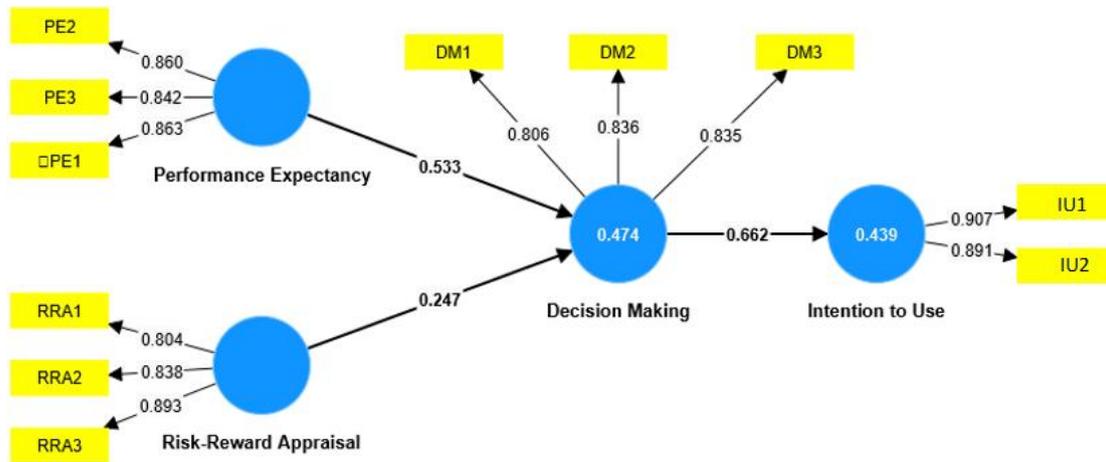

Figure 2. Measurement model

Table 2. Summary of the results of the assessments

| Construct | Item | FL | α | Composite reliability | AVE | $R^2$ |
|---|---|---|---|---|---|---|
| PE | PE1 | 0.863 | 0.816 | 0.817 | 0.731 | |
| | PE2 | 0.860 | | | | |
| | PE3 | 0.842 | | | | |
| RRA | RRA1 | 0.804 | 0.801 | 0.818 | 0.715 | |
| | RRA2 | 0.838 | | | | |
| | RRA3 | 0.893 | | | | |
| DM | DM1 | 0.806 | 0.766 | 0.766 | 0.682 | 0.474 |
| | DM2 | 0.836 | | | | |
| | DM3 | 0.835 | | | | |
| IU | IU1 | 0.907 | 0.764 | 0.767 | 0.809 | 0.439 |
| | IU2 | 0.891 | | | | |
| Average scores | | | | | 0.734 | 0.457 |
| AVE*$R^2$ | | | | | 0.335 | |
| √AVE*$R^2$ (GoF) | | | | | 0.579 | |

Table 3. Fornell-Larcker criterion

| Construct | DM | IU | PE | RRA |
|---|---|---|---|---|
| DM | 0.826 | | | |
| IU | 0.662 | 0.899 | | |
| PE | 0.654 | 0.634 | 0.855 | |
| RRA | 0.508 | 0.480 | 0.489 | 0.846 |

The explanatory power of the predictor variables on their respective constructs is reflected in the corrected $R^2$ values in Figure 2. The two dependent constructs, DM ($R^2$=0.474) and IU ChatGPT ($R^2$=0.439), both exceed the required threshold, as indicated in Table 2. This categorizes both DM and IU ChatGPT as moderate in explanatory strength (i.e., $R^2$>0.33). The influence of the independent constructs on the dependent variables was assessed using standardized path analysis to test the hypothesized relationships. Three hypotheses (H1 to H3) had positive path coefficients and were significant at the $p<0.001$ level, as shown in Table 4 and Figure 2. Therefore, all hypotheses are supported. The heterotrait-monotrait (HTMT) ratios in Table 5 assess the distinctiveness of the constructs measured in our PLS-SEM model. While the commonly recommended threshold for good discriminant validity is 0.85, this study adopted a slightly less stringent criterion of 0.90. Based on this criterion, all HTMT values fall within the acceptable range which indicates a good discriminant validity.

This study examines the perceptions of novice programmers regarding their IU ChatGPT for programming-related tasks, based on the framework developed by Shahsavar and Choudhury [48]. The factors considered include PE, RA, DM, and the IU. The survey results supported the model, confirming all three hypotheses.

Table 4. Structural estimates: path coefficients

|    | Construct | Original sample | Standard deviation | t-statistics (jb/STDEVj) | p-values |
|----|-----------|-----------------|--------------------|--------------------------|----------|
| H1 | PE→DM     | 0.533           | 0.531              | 0.054                    | 0.00     |
| H2 | RRA→DM    | 0.247           | 0.250              | 0.051                    | 0.00     |
| H3 | DM→IU     | 0.662           | 0.665              | 0.040                    | 0.00     |

Table 5. HTMT ratio

| Construct | DM    | IU    | PE    | RRA |
|-----------|-------|-------|-------|-----|
| DM        |       |       |       |     |
| IU        | 0.864 |       |       |     |
| PE        | 0.826 | 0.801 |       |     |
| RRA       | 0.643 | 0.615 | 0.606 |     |

### 4.1. Performance expectancy and decision-making

Hypothesis 1 posited that PE is positively associated with DM. The results indicate that higher PE in ChatGPT correlates with improved DM for programming tasks. One possible reason for this result is that novice programmers might believe that ChatGPT's capabilities will encourage them to explore, learn concepts easily, and broaden their skills. For instance, they might ask ChatGPT to explain specific functions or code snippets. Schukow et al. [50] found that ChatGPT can distill and summarize vast amounts of data quickly, aiding in building a foundation of knowledge on specific topics. When facing challenges, trusting ChatGPT's assistance can help users persevere through difficulties, leading to a more positive learning experience and a greater willingness to tackle complex problems [25], [51]. Additionally, similar to how AI models supplement clinician knowledge and DM processes [52], novice programmers with high PE likely believe in ChatGPT's capabilities to assist them effectively [53], [54]. This confidence might lead to better DM as users feel more assured about the information and suggestions provided by ChatGPT. Past positive experiences with ChatGPT, where it helped users solve problems efficiently, also contribute to users relying on ChatGPT more confidently for programming tasks, thus enhancing DM. If users expect high performance from ChatGPT, they may inherently look for and recognize positive outcomes and solutions, reinforcing their DM process.

### 4.2. Risk-reward appraisal and decision-making

Hypothesis 2 examined the influence of RRA on DM. It shows that a favorable risk-reward assessment of ChatGPT is linked to better DM in programming tasks. A favorable risk-reward assessment suggests that users acknowledge potential risks (such as receiving bad advice) but believe the potential benefits (such as learning new concepts or solving problems faster) outweigh those risks. This calculated approach can lead them to experiment with the tool in a way that enhances their DM. Individuals tend to evaluate the potential outcomes of their decisions based on the balance between risk and reward, with a favorable risk-reward assessment often leading to more confident and effective DM [55]. Novice programmers who perceive a favorable risk-reward ratio with ChatGPT might likely to feel more secure in using the tool. They are more likely to decide to use ChatGPT if they perceive that the benefits (such as gaining quick solutions, learning new programming techniques, and receiving immediate feedback) outweigh the potential risks (such as receiving incorrect information or becoming overly reliant on the tool). This favorable cost-benefit analysis makes the decision to use ChatGPT more appealing. When novice programmers recognize that using ChatGPT can provide significant value, such as saving time and enhancing their learning experience, they might more likely to decide to use it. The perception of added value increases the attractiveness of the decision.

### 4.3. Decision-making and intention to use ChatGPT

Hypothesis 3 revealed that DM is positively associated with the IU ChatGPT. This implies that a positive view of ChatGPT's role in DM is connected to a greater inclination to use it for programming tasks. When users have positive experiences with ChatGPT in their DM process, it reinforces their confidence in the tool's efficacy [56]–[58]. This confidence can naturally lead to a greater inclination to use ChatGPT for future tasks. A positive view of ChatGPT's role in DM builds trust in its reliability and capabilities [59]. When users trust the tool, they are more likely to intend to use it as a regular part of their workflow [60]. Users who experience improved DM with ChatGPT perceive it as adding significant value to their work. This perceived value increases their IU the tool because it enhances their productivity and problem-solving capabilities [57]. According to the theory of planned behavior, a positive attitude towards a behavior (in this case, using ChatGPT for DM) enhances the intention to perform that behavior. If users have a favorable view of ChatGPT's impact on their decisions, they are more likely to intend to use it. In the context of novice programmers, positive DM outcomes when using ChatGPT create a feedback loop where users associate the tool with successful results. This reinforcement encourages them to rely on ChatGPT more frequently for programming tasks. The DM process regarding the utilization of ChatGPT for programming tasks is positively associated with the IU ChatGPT. This connection implies that a favorable perception of ChatGPT's role in DM is linked to a higher inclination to employ it for programming activities [61], [62].

## 5. CONCLUSION

This study used quantitative methods to investigate the novice programmers' PE, RRA, DM, and IU ChatGPT. A model was developed and empirically validated, explaining 47.4% of DM and 43.9% of the IU ChatGPT among novice programmers. The research tested three key hypotheses. Hypothesis 1 proposed that PE is positively associated with DM, and the results supported this. Hypothesis 2 explored the impact of RRA on DM, which was found to be significant. Hypothesis 3 established that DM is positively correlated with the IU ChatGPT, further validating the model. This study revealed that novice programmers are motivated to use ChatGPT. They perceived it as a beneficial and credible tool for programming-related tasks. This research highlights the potential of ChatGPT to enhance the programming experience for beginners. This study also emphasized its potential value as a reliable and advantageous resource for programming education.

## FUNDING INFORMATION

This study was funded by Don Honorio Ventura State University.

## AUTHOR CONTRIBUTIONS STATEMENT

This journal uses the Contributor Roles Taxonomy (CRediT) to recognize individual author contributions, reduce authorship disputes, and facilitate collaboration.

| Name of Author | C | M | So | Va | Fo | I | R | D | O | E | Vi | Su | P | Fu |
|---|---|---|---|---|---|---|---|---|---|---|---|---|---|---|
| John Paul P. Miranda | ✓ | ✓ |   | ✓ | ✓ | ✓ | ✓ | ✓ | ✓ | ✓ | ✓ | ✓ | ✓ | ✓ |
| Jaymark A. Yambao |   | ✓ | ✓ | ✓ | ✓ | ✓ | ✓ | ✓ | ✓ | ✓ |   |   | ✓ | ✓ |

| C | : | **C**onceptualization | I | : | **I**nvestigation | Vi | : | **Vi**sualization |
|---|---|---|---|---|---|---|---|---|
| M | : | **M**ethodology | R | : | **R**esources | Su | : | **Su**pervision |
| So | : | **So**ftware | D | : | **D**ata Curation | P | : | **P**roject administration |
| Va | : | **Va**lidation | O | : | Writing - **O**riginal Draft | Fu | : | **Fu**nding acquisition |
| Fo | : | **Fo**rmal analysis | E | : | Writing - Review & **E**diting | | | |

## CONFLICT OF INTEREST STATEMENT

Authors state no conflict of interest.

## INFORMED CONSENT

The authors obtained informed consent from all individuals included in this study.

## ETHICAL APPROVAL

Ethical conduct for this research was guided by the tenets of the Belmont Report, the Declaration of Helsinki, the Philippine Data Privacy Act of 2012, and the Philippine Council for Health Research and Development's (PCHRD) National Ethical Guidelines for Research Involving Human Participants (2022). This study was approved by the author's University Research Management Office.

## DATA AVAILABILITY

Derived data supporting the findings of this study are available from the corresponding author, [JPPM], upon request.

**REFERENCES**

[1] S. A. Fincher and A. V. Robins, "The cambridge handbook of computing education research," in *The Cambridge Handbook of Computing Education Research*, S. A. Fincher and A. V. Robins, Eds. Cambridge University Press, 2019, pp. 1–905.
[2] M. E. Dacey, "A Study of Novice Programmer Performance and Programming Pedagogy," University of Wales Trinity St David, 2018.
[3] C. Izu et al., "Fostering program comprehension in novice programmers-learning activities and learning trajectories," in *Annual Conference on Innovation and Technology in Computer Science Education, ITiCSE*, 2019, pp. 27–52, doi: 10.1145/3344429.3372501.
[4] C. Delozier and J. Shey, "Using visual programming games to study novice programmers," *International Journal of Serious Games*, vol. 10, no. 2, pp. 115–136, 2023, doi: 10.17083/ijsg.v10i2.577.
[5] U. Z. Ahmed, R. Sindhgatta, N. Srivastava, and A. Karkare, "Targeted example generation for compilation errors," in *2019 34th IEEE/ACM International Conference on Automated Software Engineering, ASE 2019*, 2019, pp. 327–338, doi: 10.1109/ASE.2019.00039.
[6] S. I. Malik, M. Al-Emran, R. Mathew, R. M. Tawafak, and G. AlFarsi, "Comparison of e-learning, m-learning and game-based learning in programming education," *International Journal of Emerging Technologies in Learning*, vol. 15, no. 15, pp. 133–146, Aug. 2020, doi: 10.3991/ijet.v15i15.14503.
[7] F. Corno, L. De Russis, and J. P. Saénz, "Easing IoT development for novice programmers through code recipes," in *Proceedings - International Conference on Software Engineering*, 2018, pp. 13–16, doi: 10.1145/3183377.3183385.
[8] S. Iqbal and O. K. Harsh, "A self review and external review model for teaching and assessing novice programmers," *International Journal of Information and Education Technology*, vol. 3, no. 2, pp. 120–123, 2013, doi: 10.7763/ijiet.2013.v3.247.
[9] F. Corno, L. de Russis, and L. Mannella, "Perception of security issues in the development of cloud-IoT systems by a novice programmer," *Intelligent Environments 2021: Workshop Proceedings of the 17th International Conference on Intelligent Environments*, vol. 29, pp. 5–15, 2021, doi: 10.3233/AISE210074.
[10] L. Laporte and B. Zaman, "Informing content-driven design of computer programming games: a problems analysis and a game review," in *ACM International Conference Proceeding Series*, 2016, pp. 1–10, doi: 10.1145/2971485.2971499.
[11] R. M. A. Alhazmi, R. Maaroufi, and A. Alhazmi, "The impact of guided metacognitive feedback on novice programmers using learning by teaching environment," *Journal of King Abdulaziz University Computing and Information Technology Sciences*, vol. 8, no. 2, pp. 13–31, 2019, doi: 10.4197/comp.8-2.2.
[12] Y. Ishii and K. Asai, "Report on a user test and extension of a type debugger for novice programmers," *Electronic Proceedings in Theoretical Computer Science, EPTCS*, vol. 170, pp. 1–18, 2014, doi: 10.4204/EPTCS.170.1.
[13] F. Corno, L. de Russis, and J. P. Saenz, "Pain points for novice programmers of ambient intelligence systems: an exploratory study," in *Proceedings - International Computer Software and Applications Conference*, 2017, vol. 1, pp. 250–255, doi: 10.1109/COMPSAC.2017.186.
[14] H. Alkaissi and S. I. McFarlane, "Artificial hallucinations in ChatGPT: implications in scientific writing," *Cureus*, vol. 15, no. 2, 2023, doi: 10.7759/cureus.35179.
[15] A. J. Rojas, "An investigation into ChatGPT's application for a scientific writing assignment," *Journal of Chemical Education*, vol. 101, no. 5, pp. 1959–1965, May 2024, doi: 10.1021/acs.jchemed.4c00034.
[16] S. García-Méndez, F. de Arriba-Pérez, and M. del C. Somoza-López, "A review on the use of large language models as virtual tutors," *Science and Education*, 2024, doi: 10.1007/s11191-024-00530-2.
[17] T. A. Rengers, C. A. Thiels, and H. Salehinejad, "Academic surgery in the era of large language models: a review," *JAMA Surgery*, vol. 159, no. 4, pp. 445–450, Apr. 2024, doi: 10.1001/jamasurg.2023.6496.
[18] L. Yan et al., "Practical and ethical challenges of large language models in education: a systematic scoping review," *British Journal of Educational Technology*, vol. 55, no. 1, pp. 90–112, Jan. 2024, doi: 10.1111/bjet.13370.
[19] R. Bringula, "What do academics have to say about ChatGPT? a text mining analytics on the discussions regarding ChatGPT on research writing," *AI and Ethics*, 2023, doi: 10.1007/s43681-023-00354-w.
[20] T. Trust, J. Whalen, and C. Mouza, "Editorial: ChatGPT: challenges, opportunities, and implications for teacher education," *Contemporary Issues in Technology and Teacher Education*, vol. 23, no. 1, pp. 1–23, 2023.
[21] S. Elbanna and L. Armstrong, "Exploring the integration of ChatGPT in education: adapting for the future," *Management and Sustainability: An Arab Review*, vol. 3, no. 1, pp. 16–29, Jan. 2024, doi: 10.1108/MSAR-03-2023-0016.
[22] I. Maita, S. Saide, A. M. Putri, and D. Muwardi, "Pros and cons of artificial intelligence-ChatGPT adoption in education settings: a literature review and future research agendas," *IEEE Engineering Management Review*, vol. 52, no. 3, pp. 27–42, 2024, doi: 10.1109/EMR.2024.3394540.
[23] S. A. Bin-Nashwan, M. Sadallah, and M. Bouteraa, "Use of ChatGPT in academia: academic integrity hangs in the balance," *Technology in Society*, vol. 75, 2023, doi: 10.1016/j.techsoc.2023.102370.
[24] R. Michel-Villarreal, E. Vilalta-Perdomo, D. E. Salinas-Navarro, R. Thierry-Aguilera, and F. S. Gerardou, "Challenges and opportunities of generative AI for higher education as explained by ChatGPT," *Education Sciences*, vol. 13, no. 9. 2023, doi: 10.3390/educsci13090856.
[25] T. Rasul et al., "The role of ChatGPT in higher education: benefits, challenges, and future research directions," *Journal of Applied Learning and Teaching*, vol. 6, no. 1, pp. 41–56, 2023, doi: 10.37074/jalt.2023.6.1.29.
[26] S. R. Haghighi, M. P. Saqalaksari, and S. N. Johnson, "Artificial intelligence in ecology: a commentary on a Chatbot's perspective," *The Bulletin of the Ecological Society of America*, vol. 104, no. 4, Oct. 2023, doi: 10.1002/bes2.2097.
[27] R. Hashem, N. Ali, F. El Zein, P. Fidalgo, and O. A. Khurma, "AI to the rescue: exploring the potential of ChatGPT as a teacher ally for workload relief and burnout prevention," *Research and Practice in Technology Enhanced Learning*, vol. 19, 2024, doi: 10.58459/rptel.2024.19023.
[28] R. Bringula, "ChatGPT in a programming course: benefits and limitations," *Frontiers in Education*, vol. 9. 2024, doi: 10.3389/feduc.2024.1248705.
[29] C. A. Philbin, "Exploring the potential of artificial intelligence program generators in computer programming education for students," *ACM Inroads*, vol. 14, no. 3, pp. 30–38, Aug. 2023, doi: 10.1145/3610406.
[30] N. Nascimento, P. Alencar, and D. Cowan, "Artificial Intelligence vs. software engineers: an empirical study on performance and efficiency using ChatGPT," in *Proceedings of the 33rd Annual International Conference on Computer Science and Software Engineering*, 2023, pp. 24–33.
[31] A. Azaria, R. Azoulay, and S. Reches, "ChatGPT is a remarkable tool—for experts," *Data Intelligence*, vol. 6, no. 1, pp. 240–296, Feb. 2024, doi: 10.1162/dint_a_00235.

[32] N. Gupta, A. Rajput, and S. Chimalakonda, *COSPEX: a program comprehension tool for novice programmers*. vol. 1, no. 1. Association for Computing Machinery, 2022.
[33] C. A. G. da Silva, F. N. Ramos, R. V. de Moraes, and E. L. dos Santos, "ChatGPT: Challenges and benefits in software programming for higher education," *Sustainability (Switzerland)*, vol. 16, no. 3. 2024, doi: 10.3390/su16031245.
[34] F. V. Pantelimon and B. Ș. Posedaru, "Improving programming activities using ChatGPT: a practical approach," in *Smart Innovation, Systems and Technologies*, 2024, vol. 367, pp. 307–316, doi: 10.1007/978-981-99-6529-8_26.
[35] E. Chen, R. Huang, H.-S. Chen, Y.-H. Tseng, and L.-Y. Li, "GPTutor: A ChatGPT-powered programming tool for code explanation," in *International Conference on Artificial Intelligence in Education*, 2023, pp. 321–327.
[36] E. Shue, L. Liu, B. Li, Z. Feng, X. Li, and G. Hu, "Empowering beginners in bioinformatics with ChatGPT," *Quantitative Biology*, vol. 11, no. 2, pp. 105–108, Jun. 2023, doi: 10.15302/J-QB-023-0327.
[37] M. Wieser, K. Schöffmann, D. Stefanics, A. Bollin, and S. Pasterk, "Investigating the role of ChatGPT in supporting text-based programming education for students and teachers," in *Informatics in Schools. Beyond Bits and Bytes: Nurturing Informatics Intelligence in Education*, 2023, pp. 40–53, doi: 10.1007/978-3-031-44900-0_4.
[38] S. Speth, N. Meisner, and S. Becker, "Investigating the use of AI-generated exercises for beginner and intermediate programming courses: a ChatGPT case study," in *Software Engineering Education Conference, Proceedings*, 2023, pp. 142–146, doi: 10.1109/CSEET58097.2023.00030.
[39] A. Zuiderwijk, M. Janssen, and Y. K. Dwivedi, "Acceptance and use predictors of open data technologies: drawing upon the unified theory of acceptance and use of technology," *Government Information Quarterly*, vol. 32, no. 4, pp. 429–440, 2015, doi: 10.1016/j.giq.2015.09.005.
[40] L. M. Maruping, H. Bala, V. Venkatesh, and S. A. Brown, "Going beyond intention: integrating behavioral expectation into the unified theory of acceptance and use of technology," *Journal of the Association for Information Science and Technology*, vol. 68, no. 3, pp. 623–637, Mar. 2017, doi: 10.1002/asi.23699.
[41] V. Venkatesh, J. Y. L. Thong, and X. Xu, "Consumer acceptance and use of information technology: extending the unified theory of acceptance and use of technology," *MIS Quarterly: Management Information Systems*, vol. 36, no. 1, pp. 157–178, May 2012, doi: 10.2307/41410412.
[42] V. Venkatesh, M. G. Morris, G. B. Davis, and F. D. Davis, "User acceptance of information technology: toward a unified view," *MIS Quarterly: Management Information Systems*, vol. 27, no. 3, pp. 425–478, May 2003, doi: 10.2307/30036540.
[43] S. A. Sair and R. Q. Danish, "Effect of performance expectancy and effort expectancy on the mobile commerce adoption intention through personal innovativeness among Pakistani consumers," *Pakistan Journal of Commerce and Social Sciences*, vol. 12, no. 2, pp. 501–520, 2018.
[44] U. A. Bukar, M. S. Sayeed, S. F. A. Razak, S. Yogarayan, and O. A. Amodu, "An integrative decision-making framework to guide policies on regulating ChatGPT usage," *PeerJ Computer Science*, vol. 10, 2024, doi: 10.7717/peerj-cs.1845.
[45] S. Chavanayarn, "Navigating ethical complexities through epistemological analysis of ChatGPT," *Bulletin of Science, Technology and Society*, vol. 43, no. 3–4, pp. 105–114, Nov. 2023, doi: 10.1177/02704676231216355.
[46] X. Xu, X. Wang, Y. Zhang, and R. Zheng, "Applying ChatGPT to tackle the side effects of personal learning environments from learner and learning perspective: An interview of experts in higher education," *PLOS ONE*, vol. 19, no. 1, Jan. 2024.
[47] A. Tlili *et al.*, "What if the devil is my guardian angel: ChatGPT as a case study of using chatbots in education," *Smart Learning Environments*, vol. 10, no. 1, 2023, doi: 10.1186/s40561-023-00237-x.
[48] Y. Shahsavar and A. Choudhury, "User intentions to use ChatGPT for self-diagnosis and health-related purposes: cross-sectional survey study," *JMIR Human Factors*, vol. 10, 2023, doi: 10.2196/47564.
[49] R. Bringula, G. M. Manabat, M. A. Tolentino, and E. Torres, "Predictors of errors of novice java programmers," *World Journal of Education*, vol. 2, no. 1, pp. 3–15, 2012, doi: 10.5430/wje.v2n1p3.
[50] C. Schukow *et al.*, "Application of ChatGPT in routine diagnostic pathology: promises, pitfalls, and potential future directions," *Advances in Anatomic Pathology*, vol. 31, no. 1, pp. 15–21, 2024, doi: 10.1097/PAP.0000000000000406.
[51] S. Sharma and R. Yadav, "Chat GPT–a technological remedy or challenge for education system," *Global Journal of Enterprise Information System*, vol. 14, no. 4, pp. 46–51, May 2023.
[52] E. Goh *et al.*, "ChatGPT influence on medical decision-making, bias, and equity: a randomized study of clinicians evaluating clinical vignettes," *medRxiv*, Jan. 2023, doi: 10.1101/2023.11.24.23298844.
[53] X. Ma and Y. Huo, "Are users willing to embrace ChatGPT? exploring the factors on the acceptance of chatbots from the perspective of AIDUA framework," *Technology in Society*, vol. 75, 2023, doi: 10.1016/j.techsoc.2023.102362.
[54] Y. Zhang, Q. V. Liao, and R. K. E. Bellamy, "Efect of confidence and explanation on accuracy and trust calibration in AI-assisted decision making," in *Proceedings of the 2020 Conference on Fairness, Accountability, and Transparency*, 2020, pp. 295–305, doi: 10.1145/3351095.3372852.
[55] F. Li, H. Deng, K. Leung, and Y. Zhao, "Is perceived creativity-reward contingency good for creativity? the role of challenge and threat appraisals," *Human Resource Management*, vol. 56, no. 4, pp. 693–709, Jul. 2017, doi: 10.1002/hrm.21795.
[56] T. H. Kung *et al.*, "Performance of ChatGPT on USMLE: potential for AI-assisted medical education using large language models," *PLOS Digital Health*, vol. 2, no. 2, Feb. 2023, doi: 10.1371/journal.pdig.0000198.
[57] R. Rajjoub *et al.*, "ChatGPT and its role in the decision-making for the diagnosis and treatment of lumbar spinal stenosis: a comparative analysis and narrative review," *Global Spine Journal*, vol. 14, no. 3, pp. 998–1017, Aug. 2024, doi: 10.1177/21925682231195783.
[58] A. Rao *et al.*, "Assessing the utility of ChatGPT throughout the entire clinical workflow," *medRxiv : the preprint server for health sciences*, Jan. 2023, doi: 10.1101/2023.02.21.23285886.
[59] A. Choudhury and H. Shamszare, "Investigating the impact of user trust on the adoption and use of ChatGPT: survey analysis," *Journal of Medical Internet Research*, vol. 25, 2023, doi: 10.2196/47184.
[60] H. C. Temperley *et al.*, "Current applications and future potential of ChatGPT in radiology: a systematic review," *Journal of Medical Imaging and Radiation Oncology*, vol. 68, no. 3, pp. 257–264, Apr. 2024, doi: 10.1111/1754-9485.13621.
[61] B. Foroughi *et al.*, "Determinants of intention to use ChatGPT for educational purposes: findings from PLS-SEM and fsQCA," *International Journal of Human-Computer Interaction*, vol. 40, no. 17, pp. 4501–4520, 2024, doi: 10.1080/10447318.2023.2226495.
[62] J. Kumar, M. Rani, G. Rani, and V. Rani, "Human-machine dialogues unveiled: an in-depth exploration of individual attitudes and adoption patterns toward AI-powered ChatGPT systems," *Digital Policy, Regulation and Governance*, vol. 26, no. 4, pp. 435–449, Jan. 2024, doi: 10.1108/DPRG-11-2023-0167.

## BIOGRAPHIES OF AUTHORS

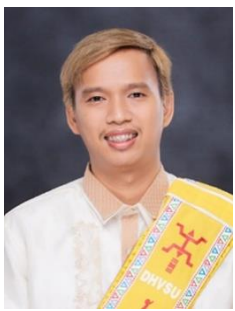 **John Paul P. Miranda** is an associate professor and the international international student scholarships, exchanges, and mobility project head for the office for international partnerships and programs at Don Honorio Ventura State University. His area of interests in publications are related to data science, analytics, educational technology, and software development. He can be contacted at email: jppmiranda@dhvsu.edu.ph.

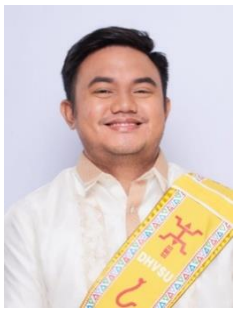 **Jaymark A. Yambao** is the program coordinator for bachelor of science in information technology at Don Honorio Ventura State University (DHVSU)-Mexico Campus. He has a master's degree in IT at System Plus College Foundation in Angeles City, Philippines. He is now pursuing his doctor of IT in the same institution with a focus on software engineering and data science. He is a member of the Philippine Society of Information Technology Educators (PSITE)-region 3 and the International Congress of Innovation-Based Educators and Researchers, Inc. He also values his personal life. His research interest includes web development, data science, and information technology education. He can be contacted at email: jayambao@dhvsu.edu.ph.